\documentclass[12pt]{iopart}

\usepackage{iopams}
\usepackage{graphicx}
\usepackage{enumerate}
\usepackage{setstack}
\usepackage{dsfont}
\usepackage{tikz}
\newcommand\tikzmark[1]{\tikz[remember picture,overlay,baseline=-1.5ex] \node (#1) {};}

\newcommand{\R}{\mathbb{R}}
\newcommand{\C}{\mathbb{C}}
\newcommand{\Z}{\mathbb{Z}}
\newcommand{\Zt}{\mathbb{Z}_{2}}
\newcommand{\Zs}{\mathbb{Z}\times\mathbb{Z}}
\newcommand{\ev}{\mathrm{ev}}
\newcommand{\od}{\mathrm{od}}
\newcommand{\GL}{\mathrm{GL}}
\newcommand{\Real}{\mathop{\mathrm{Re}}\nolimits}

\newcommand{\ad}{\mathop{\mathrm{ad}}\nolimits}
\newcommand{\Aut}{\mathop{\mathrm{Aut}}\nolimits}
\newcommand{\SAut}{\mathop{\mathrm{SAut}}\nolimits}
\newcommand{\InAut}{\mathop{\mathrm{InAut}}\nolimits}

\newcommand{\Ann}{\mathop{\mathrm{Ann}}\nolimits}
\newcommand{\1}{\mathds{1}}
\newcommand{\I}{\mathrm{i}}
\newcommand{\mybdot}{{}_{{}_{{}^{\bullet}}}}

\usepackage{lineno}

\begin{document}

\title[Unification mechanism for gauge and spacetime symmetries]{Unification mechanism for gauge and spacetime symmetries}

\author{Andr\'as L\'aszl\'o}

\address{Wigner Research Centre for Physics of the Hungarian Academy of Sciences,\\
P.O.Box 49 H-1525 Budapest, Hungary}

\ead{laszlo.andras@wigner.mta.hu}

\begin{abstract}
A group theoretical mechanism for unification of local gauge and spacetime symmetries is
introduced. No-go theorems prohibiting such unification are circumvented
by slightly relaxing the usual requirement on the gauge group: only
the so called Levi factor of the gauge group needs to be compact semisimple,
not the entire gauge group. This allows a non-conventional supersymmetry-like
extension of the gauge group, glueing together the gauge and spacetime
symmetries, but not needing any new exotic gauge particles.
It is shown that this new relaxed requirement on the gauge group
is nothing but the minimal condition for energy positivity.
The mechanism is demonstrated to be mathematically possible
and physically plausible on a $\mathrm{U}(1)$ based gauge theory setting.
The unified group, being an extension of the group of spacetime symmetries,
is shown to be different than that of the conventional supersymmetry group,
thus overcoming the McGlinn and Coleman-Mandula no-go theorems in a non-supersymmetric way.
\end{abstract}



\maketitle

\section{Introduction}
\label{secintroduction}

Unification attempts of internal (gauge) and 
spacetime symmetries is a long pursued subject in particle field theory. 
If such unification exists, it would relate coupling factors in the Lagrangian 
to each-other, which is a strong theoretical motivation. 
The non-trivialness of the problematics of such unification, however, is well-known. 
The Coleman-Mandula no-go theorem \cite{ColemanMandula1967} 
forbids the most simple unification scenarios. Namely, any larger 
symmetry group, satisfying a set of plausible properties required by a 
particle field theory context, and containing the group of spacetime symmetries 
as a subgroup as well as a gauge group, must be of the 
trivial form: gauge group $\times$ group of spacetime 
symmetries\footnote{Whenever a particle field theory model is considered on a fixed flat 
background spacetime, i.e.\ not considered as coupled to General Relativity (GR), then 
the group of spacetime symmetries is simply the Poincar\'e group. On the other hand, 
whenever a fully general relativistic field theory is studied, the group of spacetime symmetries is the 
full diffeomorphism group of the spacetime manifold, acting on the field configurations. Eventually, a general relativistic 
field theory might be also conformally invariant, in which case the group of spacetime symmetries 
is the diffeomorphism group along with conformal rescalings (Weyl rescalings) of the spacetime metric tensor field.}. 
Also, the earlier theorem of McGlinn \cite{McGlinn1964} concluded in the same direction. The classification 
result of O'Raifeartaigh \cite{LOR1965} on Poincar\'e group extensions is also usually interpreted in a similar manner. 
After the discovery of these results, the simple unification attempts of 
gauge symmetries with spacetime symmetries were not pursued further. 
Instead, a large amount of research was carried out along the question: can the Poincar\'e Lie 
algebra be extended at all in at least by means of some mathematically generalized manner? 
The answer was positive, as stated by the result of Haag, Lopuszanski and 
Sohnius \cite{HLS1975}, and hence the era of supersymmetry (SUSY) was born.

By studying the details of the proof of McGlinn and Coleman-Mandula theorems 
\cite{Weinberg2000} one finds that the assumption of the presence of a positive 
definite non-degenerate invariant scalar product on the Lie algebra of the 
gauge group is essential. Equivalently, these no-go theorems assume that the gauge 
group is of the form $\mathrm{U}(1)$ $\times$ $\dots$ $\times$ $\mathrm{U}(1)$ $\times$ 
a semisimple compact Lie group. The motivations behind this requirement are threefold:
\begin{enumerate}[(i)]
\item Group theoretical convenience: the classification 
of semisimple Lie groups is well understood.
\item Experimental justification: the 
Standard Model (SM) has a gauge group of the form $\mathrm{U}(1)\times\mathrm{SU}(2)\times\mathrm{SU}(3)$, 
which satisfies the requirement.
\item Positive energy condition: the energy density expression of a Yang-Mills (gauge) field involves the pertinent invariant 
scalar product on the Lie algebra of the gauge group, and that is required to be positive definite.
\end{enumerate}

Traditionally, gauge groups not obeying the above rule are believed to 
violate positive energy condition, and therefore are considered to be unphysical. 
However, looking more carefully, the positive energy condition merely requires 
that the invariant scalar product on the Lie algebra of the gauge group 
must be positive \emph{semidefinite}. In this paper we construct an example 
when this relaxed condition is considered, and show that this case 
is mathematically possible, physically plausible, and can 
be a key to unification of gauge and spacetime symmetries. The proposed 
mechanism can serve as an alternative to (extended) SUSY.

The structure of the paper is as follows. In Section~\ref{secliegroups} the 
Levi decomposition of Lie groups and Lie algebras are recalled, along with 
O'Raifeartaigh theorem and the elements of SUSY. In Section~\ref{secunification} 
the proposed structure for a unified gauge--Poincar\'e group is presented, which 
survives the previously recalled group theoretical constaints. In 
Section~\ref{secu1example} a concrete example group is presented for such unification, 
with $\mathrm{U}(1)$ being the compact gauge group component. 
In Section~\ref{secextendedsusy} our construction is compared to the mechanism 
of SUSY or extended SUSY. In Section~\ref{secconclusions} 
a conclusion is presented. The paper is closed by a set of Appendices, which 
expose further technical details on the concrete $\mathrm{U}(1)$ based example group.

\section{Structure of Lie groups and supersymmetry}
\label{secliegroups}

\subsection{Levi decomposition theorem}
\label{seclevi}

Recall that the symmetry group of flat spacetime, the Poincar\'e group $\mathcal{P}$ 
is composed of the group of spacetime translations $\mathcal{T}$ and of the 
homogeneous Lorentz group $\mathcal{L}$. Moreover, the group of spacetime 
translations $\mathcal{T}$ form a \emph{normal subgroup}\footnote{A subgroup 
$N$ within a larger group is called normal subgroup whenever it is invariant 
to the adjoint action of the larger group, i.e.\ whenever one has $g\,N\,g^{-1}\subset N$ for 
all elements $g$ of the larger group.} within the Poincar\'e group $\mathcal{P}$. 
Also recall that the Poincar\'e group can be written as 
$\mathcal{P}=\mathcal{T}\rtimes\mathcal{L}$, where $\rtimes$ denotes 
\emph{semi-direct product}\footnote{Semi-direct 
product means that any element of the larger group can uniquely be written 
as a product of elements from the coefficient groups, and that at least the leftmost 
coefficient group is normal subgroup. The two coefficient groups are not 
required to commute. When they commute, then also the rightmost coefficient group 
is normal subgroup, and the semi-direct product becomes a direct product, denoted by $\times$.}. 
It is seen that in the above formula $\mathcal{T}$ is an abelian normal subgroup of $\mathcal{P}$, 
and that the subgroup $\mathcal{L}$ of $\mathcal{P}$ is a simple matrix group. 
The Levi decomposition theorem \cite{IseTakeuchi1991} states that such 
decomposition property is generic to all Lie groups. Namely, any 
Lie group, assumed now to be connected and simply connected for simplicity, 
has the structure $R\rtimes L$, $R$ being a solvable 
normal subgroup called the \emph{radical} and $L$ being a semisimple 
subgroup called the \emph{Levi factor}.
The \emph{semisimpleness} of $L$ means that the \emph{Killing form} 
$(x,y)\mapsto \Tr(\ad_{x}\ad_{y})$ is non-degenerate on the Lie algebra of 
$L$, using the symbol $\ad_{x}(\cdot):=[x,\cdot]$ for any Lie algebra 
element $x$. The \emph{solvability} of $R$ means that it represents the 
degenerate directions of the Killing form. It may also be formulated in terms 
of an equivalent property: for the Lie algebra 
$r$ of $R$ with the definition 
$r^{0}:=r$, $r^{1}:=[r^{0},r^{0}]$, $r^{2}:=[r^{1},r^{1}]$, \dots, $r^{k}:=[r^{k-1},r^{k-1}]$, \dots, 
one has $r^{k}=\{0\}$ for finite $k$. 
A special case is when the radical $R$ is said to be \emph{nilpotent}: 
there exists a finite $k$ for which for all $x_{1},\dots,x_{k}\in r$ one has 
$\ad_{x_{1}}\dots\ad_{x_{k}}=0$. 
An even more special case is when the radical $R$ is \emph{abelian}: for all 
$x\in r$, one has $\ad_{x}=0$.

The (proper) Poincar\'e group with its structure $\mathcal{T}\rtimes\mathcal{L}$ 
is a demonstration of Levi decomposition theorem, where $\mathcal{T}$ is 
the abelian normal subgroup consisting of spacetime translations, being the 
radical, and where $\mathcal{L}$ is the semisimple subgroup consisting of the 
(proper) homogeneous Lorentz transformations, being the Levi factor. Groups 
like $\mathrm{SU}(N)$, often turning up as gauge groups in Yang-Mills models, 
however are semisimple, and therefore their radical vanishes, i.e.\ such a 
group consists purely of its Levi factor. Historically, groups with 
nonvanishing radical are usually not studied in context with physical field 
theory models, even though the symmetry group of flat spacetime readily provides 
an archetypical example for such groups.

\subsection{Levi structure of supersymmetry group}
\label{secsusy}

\begin{figure}
\begin{center}
\includegraphics[width=\textwidth]{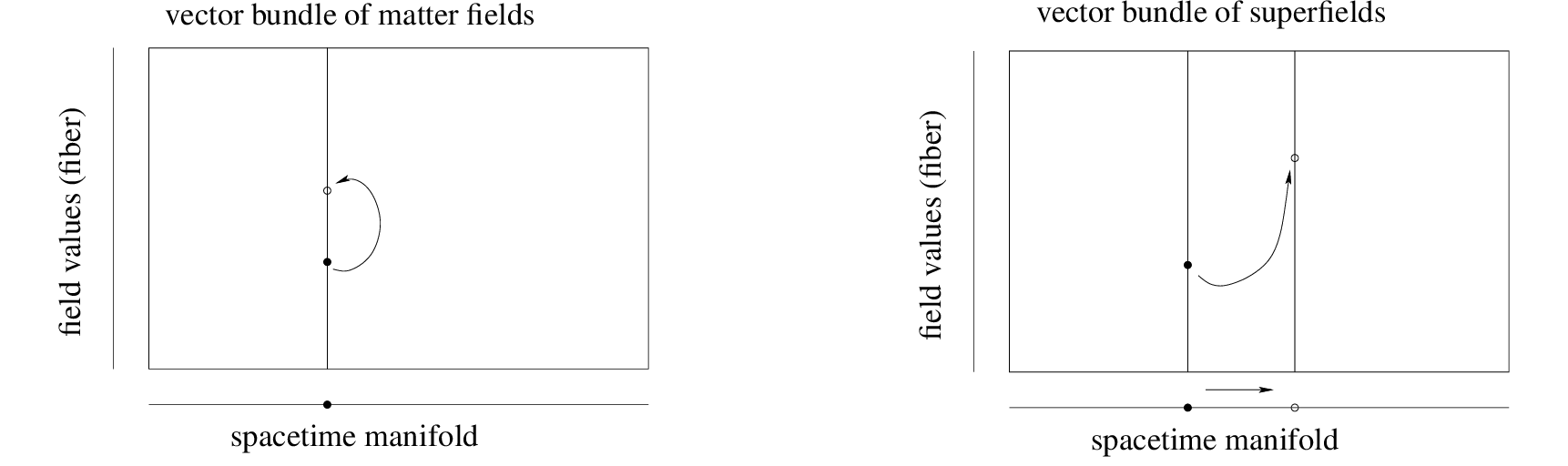}
\end{center}
\caption{Left panel: Illustration of how in a conventional gauge theory the 
gauge symmetries, i.e.\ the transformations complementing the spacetime symmetries 
act on the vector bundle of matter fields. The action of the such transformations do preserve the spacetime points, 
i.e.\ they act ``pointwise'' on the matter fields. Our construction for a unified 
gauge -- spacetime symmetry group shall also possess such property. 
Right panel: Illustration of how in a supersymmetric theory the transformations 
complementing the spacetime symmetries (i.e.\ the pure supertranslations) act 
on the vector bundle of superfields. Such a transformation does not act 
``pointwise'', but maps a field value into a field value over a shifted point of spacetime. 
The amount of shift depends also on the field value, and therefore cannot be compensated by a counter-translation.}
\label{figsusy}
\end{figure}

The Levi decomposition theorem also sheds a light 
on the group structure of supersymmetry transformations, being an extension of the Poincar\'e group. That Lie group has a Levi 
decomposition of the form $\mathcal{S}\rtimes\mathcal{L}$, where 
$\mathcal{S}$ is the nilpotent normal subgroup consisting of \emph{supertranslations}, being the radical, 
and where $\mathcal{L}$ is the semisimple subgroup consisting of the (proper) homogeneous 
Lorentz transformations, being the Levi factor. The supertranslations 
are defined as transformations on the vector bundle of superfields \cite{SS1974,FZW1974,Ferrara1987}. 
With supertranslation parameters $(\epsilon^{A},\,d^{a})$ they are of the form
\begin{eqnarray}
\left(
\begin{array}{l}
\theta^{A}\\
x^{a}\\
\end{array}
\right)
 & \mapsto &
\left(
\begin{array}{l}
\theta^{A} + \epsilon^{A}\\
x^{a} + d^{a} + \sigma^{a}_{AA'}\I\big(\theta^{A}\bar{\epsilon}^{A'}-\epsilon^{A}\bar{\theta}^{A'}\big)\\
\end{array}
\right)
\label{eqsupertr}
\end{eqnarray}
in terms of ``supercoordinates'' (Grassmann valued two-spinors) and affine spacetime 
coordinates.\footnote{A note about the presentation of supersymmetry transformations: usually, 
they are presented in the infinitesimal form and in a parametrization which 
is often referred to as a ``graded Lie algebra'', or ``super Lie algebra''. That form, however, may be 
reparametrized in order to form a conventional Lie algebra, as shown in 
\cite{SS1974,FZW1974,Ferrara1987}, see also Section~\ref{secextendedsusy}. This Lie algebra presentation, when exponentiated, shall 
form a conventional Lie group discussed above. This simple reparametrization, 
although is known in the literature \cite{SS1974,FZW1974,Ferrara1987}, is mostly not used 
in the traditional way of presentation of SUSY.}
From Eq.(\ref{eqsupertr}) it is seen that although the pure spacetime translations $\mathcal{T}$ 
form an abelian normal subgroup inside $\mathcal{S}$, but $\mathcal{S}\neq\mathcal{T}\rtimes\{\mathrm{some}\;\mathrm{other}\;\mathrm{subgroup}\}$, 
and thus such splitting is not applicable for the entire supersymmetry group. 
A geometric consequence of that phenomenon is illustrated in Figure~\ref{figsusy}: 
a pure supertranslation with parameter $(\epsilon^{A},0)$ does not act \emph{pointwise} (or \emph{fibrewise}), 
but it transforms a superfield value at a point of spacetime to an other superfield 
value over a point shifted by a corresponding spacetime translation. Note that 
such shift cannot be compensated by a counter-translation, because the introduced 
spacetime point shift depends on the field value in the fiber, i.e.\ is not 
constant as a function of the supercoordinate.

In this paper, however, we shall present a different nontrivial Poincar\'e group extension, 
enlarged both on the side of the radical and of the Levi factor, containing 
both the gauge and the spacetime symmetries, and being of the form
\begin{eqnarray}
\mathcal{T}\rtimes\{\mathrm{some\;group\;acting\;at\;points\;of\;spacetime}\},
\label{eqlocalglobal}
\end{eqnarray}
and thus rather acting pointwise, similarly as conventional gauge groups do, 
as illustrated in the left panel of Figure~\ref{figsusy}.

\subsection{O'Raifeartaigh classification of Poincar\'e group extensions}
\label{seclor}

Let us take a larger symmetry group $E$ with its Levi decomposition $E=R\rtimes L$, 
containing the Poincar\'e group $\mathcal{P}=\mathcal{T}\rtimes\mathcal{L}$ as a subgroup. 
Then the theorem of O'Raifeartaigh \cite{LOR1965} states that either one has 
$\mathcal{T}\subset R$ and $\mathcal{L}\subset L$ (radical embedded into radical, 
Levi factor embedded into Levi factor), or one has $\mathcal{T}\rtimes\mathcal{L}\subset L$ 
(the entire Poincar\'e group is embedded into the Levi factor of a much 
larger symmetry group). This result leads to the following classification theorem of 
O'Raifeartaigh \cite{LOR1965} on the possible extensions of the Poincar\'e group:
\begin{enumerate}[(i)]
\item $R=\mathcal{T}$, and $L=\left\{\mathrm{some\;semisimple\;Lie\;group}\right\}\times \mathcal{L}$. 
This means that whenever the radical $R$ of the larger symmetry group solely 
consists of the spacetime translations, then one has only the trivial group extension $E=\mathcal{P}\times\left\{\mathrm{some\;extra\;symmetries}\right\}$. 
This group theoretical phenomenon drives the no-go theorems of McGlinn and Coleman-Mandula.
\item $R$ is an abelian extension of $\mathcal{T}$, and $\mathcal{L}\subset L$. 
This means that in the radical $R$ of the larger symmetry group one has the 
spacetime translations and some abelian extension. 
The Levi factor $L$ of the extended symmetries $E$ may be larger than $\mathcal{L}$.
\item $R$ is a non-abelian extension of $\mathcal{T}$, and $\mathcal{L}\subset L$. In this case 
the radical $R$ contains the spacetime translations and some non-abelian solvable extension. 
The Levi factor $L$ of the extended symmetries $E$ can be larger than $\mathcal{L}$. 
SUSY, extended SUSY, as well as the example to be presented in this paper falls into this case.
\item $\mathcal{T}\rtimes\mathcal{L}\subset L$ and $L$ is a simple Lie group. This 
case means that the entire Poincar\'e group is fully embedded into a much larger 
simple Lie group. Conformal theories, i.e.\ theories having the conformal 
Poincar\'e transformations as symmetry group\footnote{
Conformal Poincar\'e group is isomorphic to $\mathrm{SO}(2,4)$, hence it is a simple Lie group.} are typical examples. 
Also an $\mathrm{SO}(1,13)$ based theory \cite{Chamseddine2016}, as well as an 
$\mathrm{E}_{8}$ based theory \cite{Lisi2010} provide such examples. All 
of these models do need a symmetry breaking to explain a 
Standard Model-like limit of the corresponding theory, since the embedding 
group is rather large.
\end{enumerate}
Consequently: for nontrivially extending the Poincar\'e group, its radical must necessarily be extended, as shown by cases (ii)--(iii), 
or the extended group must be a spontaneously broken large simple Lie group, 
as shown for the case (iv).

It is seen that the supersymmetry group is of type (iii) in the classification 
theorem of O'Raifeartaigh: its radical is extended and therefore the no-go 
theorems of McGlinn and Coleman-Mandula are not applicable. The unification mechanism 
for gauge and spacetime symmetries proposed in the followings uses the same group 
theoretical possibility as well, but in a very different way in comparison to SUSY: 
our extended group shall have the structure Eq.(\ref{eqlocalglobal}), 
which is not the case for the SUSY group.

\section{Unification mechanism for gauge and spacetime symmetries}
\label{secunification}

In terms of global symmetries, our proposed unification mechanism for 
gauge and spacetime symmetries assumes a structure
\newline
\begin{minipage}{\textwidth}
\begin{eqnarray}
 \cr
 \cr
\underbrace{\Big(\underbrace{\overset{\tikzmark{a}}{\mathcal{T}}}_{\mathrm{translations}}\;\times\;\underbrace{\underbrace{\overset{\tikzmark{c}}{\mathcal{N}}}_{\mathrm{solvable\;internal}}\Big) \;\rtimes\; \Big(\underbrace{\overset{\tikzmark{d}}{\mathcal{G}}}_{\mathrm{compact\;internal}}}_{\mathrm{full\;gauge\;(internal)\;group}}\;\times\;\underbrace{\overset{\tikzmark{e}\;\;\tikzmark{b}}{\mathcal{L}}}_{\mathrm{Lorentz\;(or\;Weyl)\;group}}\Big)}_{\mathrm{global\;symmetries\;of\;matter\;fields\;when\;considered\;over\;flat\;spacetime}}\cr
\tikz[remember picture,overlay] \draw[<-] (a.north) |- +(0ex,4.5ex) -| (b.north);
\tikz[remember picture,overlay] \draw[<-] (c.north) |- +(0ex,2.5ex) -| (d.north);
\tikz[remember picture,overlay] \draw[<-] (c.north) |- +(0ex,2.5ex) -| (e.north);
\label{equnifiedglobal}
\end{eqnarray}
\end{minipage}
\newline
for the unified group. Here, 
$\mathcal{G}$ symbolizes the usual compact gauge group, being $\mathrm{U}(1)\times\mathrm{SU}(2)\times\mathrm{SU}(3)$ 
in case of Standard Model, $\mathcal{L}$ denotes the homogeneous part of the spacetime symmetry group, 
being the homogeneous Lorentz (or possibly, the Weyl\footnote{Weyl group: the homogeneous Lorentz group augmented by the group of metric rescalings with a constant conformal factor.}) group, and $\mathcal{N}$ 
stands for a non-usual extension of the group of internal symmetries, allowed to 
be a solvable normal subgroup.
The arrows indicate which subgroup acts nontrivially 
on which normal subgroup, i.e.\ subgroups not connected by arrows do commute, 
whereas the others do not. 
Clearly, such group structure as a Poincar\'e group extension is potentially 
allowed by the case (iii) of O'Raifeartaigh classification theorem. Using the 
semi-associativity of $\rtimes$ and $\times$, the global unified group described 
by Eq.(\ref{equnifiedglobal}) can be rewritten in an equivalent form
\newline
\begin{minipage}{\textwidth}
\begin{eqnarray}
 \cr
 \cr
\underbrace{\underbrace{\overset{\tikzmark{a}}{\mathcal{T}}}_{\mathrm{translations}} \;\rtimes\; \underbrace{\Bigg(\underbrace{\underbrace{\overset{\tikzmark{c}}{\mathcal{N}}}_{\mathrm{solvable\;internal}}\;\rtimes\;\Big(\underbrace{\overset{\tikzmark{d}}{\mathcal{G}}}_{\mathrm{compact\;internal}}}_{\mathrm{full\;gauge\;(internal)\;group}}\;\times\;\underbrace{\overset{\tikzmark{e}\;\;\tikzmark{b}}{\mathcal{L}}}_{\mathrm{Lorentz\;(or\;Weyl)\;group}}\Big)\Bigg)}_{\mathrm{symmetries\;of\;matter\;fields\;at\;points\;of\;spacetime}}}_{\mathrm{global\;symmetries\;of\;matter\;fields\;when\;considered\;over\;flat\;spacetime}}\cr
\tikz[remember picture,overlay] \draw[<-] (a.north) |- +(0ex,4.5ex) -| (b.north);
\tikz[remember picture,overlay] \draw[<-] (c.north) |- +(0ex,2.5ex) -| (d.north);
\tikz[remember picture,overlay] \draw[<-] (c.north) |- +(0ex,2.5ex) -| (e.north);
\label{equnifiedglobal2}
\end{eqnarray}
\end{minipage}
\newline
which shows that our unified group, as global symmetries, are of the form of 
Eq.(\ref{eqlocalglobal}). That naturally motivates to search for a local unified 
group of gauge and spacetime symmetries in the form
\newline
\begin{minipage}{\textwidth}
\begin{eqnarray}
\cr
\underbrace{\underbrace{\underbrace{\overset{\tikzmark{c}}{\mathcal{N}}}_{\mathrm{solvable\;internal}} \;\;\rtimes\;\; \Big(\underbrace{\overset{\tikzmark{d}}{\mathcal{G}}}_{\mathrm{compact\;internal}}}_{\mathrm{full\;gauge\;(internal)\;group}}\;\times\;\underbrace{\overset{\tikzmark{e}}{\mathcal{L}}}_{\mathrm{Lorentz\;(or\;Weyl)\;group}}\Big)}_{\mathrm{local\;symmetries\;of\;matter\;fields\;at\;points\;of\;spacetime}}\cr
\tikz[remember picture,overlay] \draw[<-] (c.north) |- +(0,2.5ex) -| (d.north);
\tikz[remember picture,overlay] \draw[<-] (c.north) |- +(0,2.5ex) -| (e.north);
\label{equnificationmechanism}
\end{eqnarray}
\end{minipage}
\newline
which is just Eq.(\ref{equnifiedglobal2}) without the translations, acting as 
local symmetry on the matter fields independently at each point of a spacetime 
manifold. Again, using the semi-associativity of $\rtimes$ and $\times$, 
the local unified group described by Eq.(\ref{equnificationmechanism}) can be 
rewritten in the equivalent form
\newline
\begin{minipage}{\textwidth}
\begin{eqnarray}
\cr
\underbrace{\underbrace{\Big(\underbrace{\overset{\tikzmark{c}}{\mathcal{N}}}_{\mathrm{solvable\;internal}} \;\rtimes\; \underbrace{\overset{\tikzmark{d}}{\mathcal{G}}}_{\mathrm{compact\;internal}}\Big)}_{\mathrm{full\;gauge\;(internal)\;group}}\;\;\rtimes\;\;\underbrace{\overset{\tikzmark{e}}{\mathcal{L}}}_{\mathrm{Lorentz\;(or\;Weyl)\;group}}}_{\mathrm{local\;symmetries\;of\;matter\;fields\;at\;points\;of\;spacetime}}\cr
\tikz[remember picture,overlay] \draw[<-] (c.north) |- +(0,2.5ex) -| (d.north);
\tikz[remember picture,overlay] \draw[<-] (c.north) |- +(0,2.5ex) -| (e.north);
\label{equnificationmechanism2}
\end{eqnarray}
\end{minipage}
\newline
which implies that there exists a \emph{homomorphism}\footnote{Group homomorphism: 
a product preserving mapping from one group to another.} from the local unified group 
Eq.(\ref{equnificationmechanism2}) onto the local group of spacetime symmetries 
$\mathcal{L}$, and the kernel of that homomorphism is the local group of internal (gauge) symmetries 
$\mathcal{N}\rtimes\mathcal{G}$. 
This finding implies the following consequences:
\begin{enumerate}[(i)]
\item The full local unified group Eq.(\ref{equnificationmechanism}) has a four-vector representation through the homomorphism onto $\mathcal{L}$.
\item The group of local internal (gauge) symmetries $\mathcal{N}\rtimes\mathcal{G}$ act trivially on such four-vector representation --- hence the name: they act trivially on the spacetime vectors.
\item The full local unified group Eq.(\ref{equnificationmechanism}) acts as the Lorentz (or Weyl) group on such four-vector representation.
\item Because of the previous point, there exists a uniquely determined Lorentz metric conformal equivalence class on the four-vector representation, preserved by the local unified group Eq.(\ref{equnificationmechanism}).
\item Because of the previous point, there exists a uniquely determined Lorentz causal structure preserved by the local unified group Eq.(\ref{equnificationmechanism}).
\item Due to the presence of $\mathcal{N}$, the local unified group Eq.(\ref{equnificationmechanism}) is indecomposable, i.e.\ is not of the form of a direct product.
\end{enumerate}

In conclusion, Eq.(\ref{equnificationmechanism}) shows that the local gauge group and the group of 
local spacetime symmetries would decompose into a direct product $\mathcal{G}\times\mathcal{L}$ as dictated 
by the McGlinn and Coleman-Mandula no-go theorems, however the solvable normal subgroup $\mathcal{N}$ 
of local gauge symmetries glues them together, making the unification. With that, 
the full local gauge group shall be an extended one, $\mathcal{N}\rtimes\mathcal{G}$, as a price to pay. 
Since $\mathcal{N}$ represents the degenerate directions of the Killing form over 
the full gauge group $\mathcal{N}\rtimes\mathcal{G}$, it only adds some 
zero-energy gauge field modes to a field theoretical model having local unified symmetries as Eq.(\ref{equnificationmechanism}). 
These zero-energy gauge field modes shall also have vanishing Yang-Mills kinetic Lagrangian term, and 
therefore such unification mechanism does not cost adding new propagating 
gauge particle fields to the system. They do contribute, however, to other 
parts of the Lagrangian involving matter fields and their covariant derivatives, 
restricting the forms of possible Lagrangians compatible with the extended symmetry 
requirement. It is remarkable, that the proposed unification mechanism does 
not necessarily need a breaking of the large symmetry group, as the non-conventional 
part $\mathcal{N}$ of internal symmetries is inapparent in terms of detectable 
gauge particles. Also, one should note that the allowed more relaxed structure 
$\mathcal{N}\rtimes\mathcal{G}$ of the full gauge group means a softer 
regularity condition than traditionally required in gauge theory: only the Levi factor of the gauge group needs to be 
compact, not the entire gauge group itself. This is equivalent to the 
positive semidefiniteness of the Killing form on the gauge group, and hence is the minimal 
requirement for the non-negativity of the energy density expression of the 
Yang-Mills fields in a system with such unified symmetries.

In the coming section we shall construct a minimal version of a unified local 
symmetry group as in Eq.(\ref{equnificationmechanism}), with $\mathcal{G}=\mathrm{U}(1)$. 
There is strong indication that the same mechanism can also be performed 
for the full Standard Model gauge group, e.g.\ using the approach of \cite{Furey2015}.

\section{\boldmath Concrete example for the 
$\mathrm{U}(1)$ case}
\label{secu1example}

Our example for a local unified symmetry group having the structure like Eq.(\ref{equnificationmechanism}) with 
$\mathcal{G}=\mathrm{U}(1)$ shall be described below. It is a non-supersymmetric extension of the (proper) homogeneous 
Lorentz (or rather, of the Weyl) group. It is detailed in \cite{Laszlo2016,Laszlo2017un} and in \ref{appu1example}.

Let $A$ be a finite dimensional 
complex unital associative algebra, with its unit denoted by $\1$. Whenever $A$ is also equipped with a 
conjugate-linear involution $(\cdot)^{+}:A\rightarrow A$ such that for all 
$x,y\in A$ one has $(xy)^{+}=x^{+}y^{+}$, then it shall be called a \emph{
${}^{+}$-algebra}. Note that this notion differs from the well-known 
mathematical notion of ${}^{*}$-algebra as here the ${}^{+}$-adjoining does not 
exchange the order of products. Let now $A$ be a finite dimensional complex 
associative algebra with unit, being also ${}^{+}$-algebra, and possessing a 
minimal generator system $(e_{1},e_{2},e_{3},e_{4})$ obeying the identity
\begin{eqnarray}
e_{i}e_{j} + e_{j}e_{i} & = & 0 \quad (i,j\in\{1,2\}\;\mathrm{or}\;i,j\in\{3,4\}), \cr
e_{i}e_{j} - e_{j}e_{i} & = & 0 \quad (i\in\{1,2\}\;\mathrm{and}\;j\in\{3,4\}), \cr
                  e_{3} & = & e_{1}^{+},\cr
                  e_{4} & = & e_{2}^{+}, \cr
e_{i_{1}}e_{i_{2}} \dots e_{i_{k}} & & (1\leq i_{1}<i_{2}<\dots<i_{k}\leq 4,\; 0\leq k\leq 4) \cr
 & & \quad \mathrm{are}\;\mathrm{linearly}\;\mathrm{independent}.
\label{eqcanonicalgen}
\end{eqnarray}
Then we call $A$ \emph{spin algebra}, and we call a minimal generator system 
obeying Eq.(\ref{eqcanonicalgen}) a \emph{canonical generator system}, whereas 
the ${}^{+}$-operation is called \emph{charge conjugation}. That is, 
spin algebra is a freely generated unital complex associative algebra with four generators, 
and the generators admit two sectors within which the generators anticommute, 
whereas the two sectors commute with each-other, and are charge conjugate to 
each-other. It is easy to check that if $S^{*}$ is a complex two dimensional 
vector space (called the \emph{cospinor space}), and $\bar{S}^{*}$ is its 
complex conjugate vector space, then $\Lambda(\bar{S}^{*})\otimes\Lambda(S^{*})$ 
naturally becomes spin algebra, where $\Lambda(\cdot)$ denotes the exterior algebra of its argument. 
It is also seen that any spin algebra is isomorphic (not naturally) to this algebra, 
i.e.\ they all have the same structure, but there is a freedom in matching the 
canonical generators. Some properties of the pertinent mathematical structure 
is listed in \cite{Laszlo2016}. In terms of a formal quantum field theory (QFT) 
analogy, the spin algebra can be regarded as the creation operator algebra of 
a fermion particle with two internal degrees of freedom along with its antiparticle, 
at a fixed point of spacetime, or equivalently, at a fixed point of momentum space. 
It is important to understand, however, that in this construction the creation 
operators of antiparticles are not yet identified with the annihilation operators 
of particles, i.e.\ it is not a canonical anticommutation relation (CAR) algebra. 
As such, the spin algebra reflects the following physical picture:
\begin{enumerate}[(i)]
\item The basic ingredients of the system are particles obeying Pauli's exclusion principle.
\item These particles have finite (two) internal degrees of freedom.
\item Corresponding charge conjugate particles are present in the system.
\end{enumerate}

Our extension of the homogeneous Weyl group shall be nothing but $\Aut(A)$, 
the \emph{automorphism group} of the spin algebra $A$. That consists of those invertible 
$A\rightarrow A$ linear transformations, which preserve the algebraic product as well as 
the charge conjugation operation.

It can be shown that if the discrete symmetries are omitted, i.e.\ if the unit connected 
component of $\Aut(A)$ is considered, then it has a structure of the form
\newline
\begin{minipage}{\textwidth}
\begin{eqnarray}
\cr
  \underbrace{\underbrace{\underbrace{\overset{\tikzmark{c}}{N}}_{\mathrm{nilpotent\;internal}}\;\rtimes\;\Big(\underbrace{\overset{\tikzmark{d}}{\mathrm{U}(1)}}_{\mathrm{compact\;internal}}}_{\mathrm{full\;gauge\;(internal)\;symmetries}}\;\times\;\underbrace{\overset{\tikzmark{e}}{\mathcal{L}}}_{\mathrm{Weyl\;group}}\Big)}_{\mathrm{symmetries\;of\;}A\mathrm{-valued\;fields\;at\;a\;point\;of\;spacetime\;or\;momentum\;space}}\cr
\tikz[remember picture,overlay] \draw[<-] (c.north) |- +(0,2.5ex) -| (d.north);
\tikz[remember picture,overlay] \draw[<-] (c.north) |- +(0,2.5ex) -| (e.north);
\label{u1exampleillustr}
\end{eqnarray}
\end{minipage}
\newline
which exactly has a structure like Eq.(\ref{equnificationmechanism}). For details 
we refer to \ref{appu1example} and \cite{Laszlo2016,Laszlo2017un}. The 
nilpotent normal subgroup $N$ of internal symmetries transform a system of 
canonical generators in such a way, that it adds higher polynomials of the 
generators to pure generators, and hence they are named ``dressing transformations''. 
Note that the pertinent example group $\Aut(A)$ can also be restricted so that 
it does not contain the conformal (Weyl) dilatations, but merely the Lorentz 
group instead of the Weyl group. That is, the inclusion or exclusion of the 
conformal dilatations to the unified group is optional: both constructions 
are group theoretically possible.

The nature of our example construction shows that 
the proposed unified symmetry group can be considered as the symmetries of 
a limiting scenario in QFT, when the position (or momentum) of fields are 
fixed and only the internal degrees of freedom are allowed to behave according 
to the algebra rules of fermionic particle and antiparticle creation operators. 
In that picture, the new nilpotent symmetries $N$ can be understood to 
mix higher particle contributions to single 
particle creation operators, and this is how the mechanism bypasses 
Coleman-Mandula theorem. (Coleman-Mandula theorem implicitly assumes that 
the symmetries do map single particle creation operators to single particle 
creation operators, which is apparently violated here.)

\section{Comparison to SUSY and extended SUSY}
\label{secextendedsusy}

In this section we show in a detailed manner that a unified gauge and spacetime 
symmetry group of the form Eq.(\ref{equnifiedglobal}) is inequivalent to SUSY 
or extended SUSY, however, they are along a similar group theoretical philosophy: 
both the (extended) SUSY and our construction use the case (iii) of O'Raifeartaigh theorem. 
On the other hand, the detailed group structure of the two constructions are different, and they use slightly 
different means to bypass Coleman-Mandula theorem.

Traditionally, the SUSY algebra is presented in a \emph{graded Lie algebra} (also called \emph{super Lie algebra}) 
form, with the following generating operators:
\newline
\begin{minipage}{\textwidth}
\begin{eqnarray}
 \Sigma_{ab} & & \mathrm{(generators\;of\;Lorentz\;Lie\;algebra)},\cr
 Q_{A} \mathrm{\,\;and\;\,} \bar{Q}_{A'} & \qquad & \mathrm{(supercharges)},\cr
 P_{a} & & \mathrm{(generators\;of\;translation\;Lie\;algebra)}
\end{eqnarray}
\end{minipage}
\newline
obeying the usual super Lie algebra relations \cite{HLS1975,SS1974,FZW1974,Ferrara1987}. 
Here, conventional Penrose abstract index notation is used \cite{PR1984,Wald1984}. 
The super Lie algebra presentation of the SUSY algebra might look paradoxical at a first glance for the following 
reason. Given a set of transformations, in which subsequent application of transformations is within the set, along with 
the identity transformation as well as inverse transformation, then that collection 
of transformations automatically obey the group axioms. (This is how the group 
axioms were distilled, at the first place.) Then, if such a set of transformations 
are parametrized by some finite tuple of real parameters, and the multiplication and inverting of 
transformations are continuously differentiable operations with respect to the parameters, 
then this set of transformations will automatically obey Lie group axioms. As such, their infinitesimal 
versions, i.e.\ their derivatives with respect to the parameters around the 
unity, automatically obey the Lie algebra axioms. Therefore, if one presents a graded 
or super Lie algebra, which does not obey ordinary Lie algebra relations, 
one needs to explain that in what sense these can be considered as infinitesimal 
version of some parametric transformations. This seemingly paradoxical question can be 
resolved by recognizing that the super Lie algebra of SUSY can be re-parametrized 
to obey ordinary Lie algebra relations \cite{SS1974,FZW1974,Ferrara1987}. 
In order to show that, take a basis $\left(\epsilon^{A}{}_{(1)},\epsilon^{A}{}_{(2)}\right)$ of the 
Grassmann valued two-spinor space, and take the definitions of the following operators:
\begin{eqnarray}
 \Sigma_{ab} & & \mathrm{(generators\;of\;Lorentz\;Lie\;algebra)},\cr
 \delta_{(i)}:=\epsilon^{A}{}_{(i)}Q_{A} \mathrm{\;and\;} \bar{\delta}_{(i)}:=\bar{\epsilon}^{A'}{}_{(i)}\bar{Q}_{A'} & \; & \mathrm{(generators\;of\;pure\;supertranslations)},\cr
 \quad(i=1,2) & & \cr
 P_{a} & & \mathrm{(generators\;of\;translation\;Lie\;algebra).}\cr
       & & 
\end{eqnarray}
It is seen that the Lorentz generators span the Lorentz Lie algebra, let us denote that by 
$\ell$, the translation generators span the translation Lie algebra, let us 
denote that by $t$, whereas the pure supertranslation generators span a subspace, 
which shall be denoted by $q$. It is seen that by considering the $\delta$-s 
(variation of superfields upon an infinitesimal pure supertranslation) instead 
of $Q$-s (supercharges) as operators acting on the superfields, 
the super Lie algebra of SUSY has an equivalent ordinary Lie algebra view, due 
to the ``sign flipping trick'' by the Grassmann valued two-spinor basis. 
It is evident, by construction, that such intertwining map between the SUSY 
super Lie algebra and the corresponding ordinary Lie algebra presentation is one-to-one and onto, 
furthermore that it really intertwines between the super and the ordinary Lie bracket 
in the two presentations. (Although this ordinary Lie algebra view of 
SUSY is known in the literature \cite{SS1974,FZW1974,Ferrara1987}, it is not 
very commonly used.) Taking now the corresponding 
ordinary Lie algebra, consisting of $t\oplus q\oplus \ell$, it is seen that 
by exponentiating it one gets a corresponding Lie group, as discussed in 
Section~\ref{secsusy} and \cite{SS1974,FZW1974,Ferrara1987}. Due to the SUSY relations, one has that the sub-Lie 
algebra of translations ($t$) is a normal sub-Lie algebra, i.e.\ it is invariant to the adjoint 
action of the entire Lie algebra $t\oplus q\oplus\ell$. Also the sub-Lie algebra of supertranslations 
($s:=t\oplus q$) is a normal sub-Lie algebra. The subspace $q$, residing within 
$s$ is merely a linear subspace, not even a sub-Lie algebra, since it does not 
close without $t$ under the Lie bracket. The subspace $\ell$ is a sub-Lie algebra, 
but it is not normal, since it acts on $s=t\oplus q$ nontrivially by the adjoint action. It is important to 
note that the normal sub-Lie algebra of translations ($t$) is abelian, and that 
the quotient Lie algebra $q\equiv s/t$ (supertranslations without considering 
the spacetime translation component) is also abelian. Exactly this structure 
makes it possible to perform the ``sign flipping trick'', i.e.\ to have a super Lie 
algebra view. Let us introduce the notation $s=t\mybdot q$ for denoting the fact that 
the Lie algebra $s=t\oplus q$ is an extension of the normal sub-Lie algebra 
$t$, but its complementing subspace $q$ is not a standalone sub-Lie algebra 
(``semi-semi-direct product''). Then, the Lie algebra view of SUSY can be 
presented as:
\newline
\begin{minipage}{\textwidth}
\begin{eqnarray}
\cr
\cr
\underbrace{\underbrace{\Big(\underbrace{\overset{\tikzmark{a}}{t}}_{\mathrm{translations}}\;\mybdot\;\underbrace{\overset{\tikzmark{c}}{q}}_{\mathrm{pure\;\,supertranslations}}\Big)}_{\mathrm{all\;\,supertranslations}\;(=s)} \;\rtimes\; \underbrace{\overset{\tikzmark{e}\;\;\tikzmark{b}}{\ell}}_{\mathrm{Lorentz\;\,symmetries}}}_{\mathrm{supersymmetries\;\,of\;\,superfields}}\cr
\tikz[remember picture,overlay] \draw[<-] (a.north) |- +(0ex,4.5ex) -| (b.north);
\tikz[remember picture,overlay] \draw[<-] (c.north) |- +(0ex,2.5ex) -| (e.north);
\label{eqsusy}
\end{eqnarray}
\end{minipage}
\newline
where again the arrow diagram clarifies which sub-Lie algebra acts nontrivially 
on which part of the Lie algebra, via the adjoint action. It is seen that 
$s=t\mybdot q$ is the radical of the full supersymmetry Lie algebra $\left(t\mybdot q\right)\rtimes\ell$, 
and that $s$ is a nilpotent extension of $t$. This group theoretical structure 
is allowed by the case (iii) of the O'Raifeartaigh theorem. The so called 
extended SUSY has very similar Lie algebra structure, with merely the abelian part 
of the radical being extended by the so called \emph{central charges}, 
being $z:=\mathrm{u}(1)\times\dots\times\mathrm{u}(1)$, and the Levi factor 
being extended by the Lie algebra of a compact gauge (internal) group $g$:
\newline
\begin{minipage}{\textwidth}
\begin{eqnarray}
\cr
\cr
\underbrace{\underbrace{\Big(\big(\underbrace{\overset{\tikzmark{a}}{t}}_{\mathrm{translations}}\times\underbrace{z}_{\mathrm{central\;charges}}\big)\mybdot\underbrace{\overset{\tikzmark{c}}{q_{{}_{\mathrm{ext}}}}}_{\substack{\mathrm{pure\;extended}\cr\mathrm{supertranslations}}}\Big)}_{\mathrm{all\;\,extended\;\,supertranslations}} \rtimes \Big(\underbrace{\overset{\tikzmark{d}}{g}}_{\substack{\mathrm{compact\;internal}\cr\mathrm{symmetries}}} \times \underbrace{\overset{\tikzmark{e}\;\;\tikzmark{b}}{\ell}}_{\substack{\mathrm{Lorentz}\cr\mathrm{symmetries}}}\Big)}_{\mathrm{extended\;\,supersymmetries}}\cr
\tikz[remember picture,overlay] \draw[<-] (a.north) |- +(0ex,4.5ex) -| (b.north);
\tikz[remember picture,overlay] \draw[<-] (c.north) |- +(0ex,2.5ex) -| (d.north);
\tikz[remember picture,overlay] \draw[<-] (c.north) |- +(0ex,2.5ex) -| (e.north);
\label{eqextsusy}
\end{eqnarray}
\end{minipage}
\newline
Here, the arrow diagram explicitly shows that the glueing of the compact gauge (internal) 
symmetries and the Lorentz symmetries are possible due to their common adjoint Lie group 
action on some subspace ($q_{{}_{\mathrm{ext}}}$) of the radical. In that sense, 
the unification mechanism Eq.(\ref{equnifiedglobal}) proposed in the present paper 
is kind of similar to that of the mechanism of extended SUSY, shown in Eq.(\ref{eqextsusy}).

Given the group theoretical similarities of the (extended) SUSY 
illustrated in Eq.(\ref{eqextsusy}), and our proposed mechanism outlined in Eq.(\ref{equnifiedglobal}), 
the question naturally arises: are these constructions inequivalent? The 
answer is yes, which shall be demonstrated in multiple ways, in the closing part of this section.

Recall that a normal sub-Lie algebra within a Lie algebra is an invariant 
subspace, and as such is independent of the choice of generators (i.e.\ of a Lie algebra basis). 
Therefore, if the list of normal sub-Lie algebras within two Lie algebras cannot be 
identified to each-other, then these Lie algebras cannot be isomorphic. Using Eq.(\ref{eqextsusy}) 
and Eq.(\ref{equnifiedglobal}) one can list the normal sub-Lie algebras in the 
two constructions, and can see that they are different in number and are 
different in terms of dimensions, i.e.\ cannot be identified to each-other.

An other way to show the inequivalence of (extended) SUSY and our unification 
mechanism is to observe that our group, by construction, can be regarded of the form of Eq.(\ref{eqlocalglobal}), 
whereas (extended) SUSY cannot be transformed into that form. 
That is seen via referring again to the invariance of the normal sub-Lie algebras, 
as a consequence of which the definition of the sub-Lie algebra consisting of translations and central charges is independent of the choice of generators in (extended) SUSY. 
Clearly, its complementing subspace ($q$) does not form a standalone sub-Lie algebra, which obstructs Eq.(\ref{eqlocalglobal}). 
It can be shown that by taking a different complementing subspace ($q'$), being some mixture 
of $q$ and of the translations and central charges, this cannot be avoided. 
That is evidently seen by taking new generators $\delta_{(i)}'$ ($i=1,2$) 
being linear combinations of $\delta_{(i)}$ and of translations and central charges, 
and then by using SUSY relations. It becomes evident that the structure relations of 
$\delta_{(i)}'$ ($i=1,2$) shall be the same as of $\delta_{(i)}$, due to:
\begin{itemize}
 \item abelian nature of translations and central charges,
 \item $\delta_{(i)}$ ($i=1,2$) commute with translations and central charges.
\end{itemize}
Therefore, no complementing sub-Lie algebra $q'$ to translations and central 
charges can be found, merely a complementing sub-linear space can exist, which then 
indeed obstructs Eq.(\ref{eqlocalglobal}) to hold for the (extended) SUSY.

A further way to see the inequivalence of the proposed unification mechanism 
and of (extended) SUSY is to observe that our construction Eq.(\ref{equnifiedglobal}) 
can be regarded as $\left(\mathcal{N}\rtimes\mathcal{G}\right)\rtimes\left\{\mathrm{Poincar\acute{e}\;group}\right\}$. 
That implies the existence of a homomorphism from that group onto the Poincar\'e group. 
The (extended) SUSY does not possess such homomorphism onto the Poincar\'e group, 
since as pointed out above, the pure supertranslation generators cannot be 
collected into a normal sub-Lie algebra (not even to an ordinary sub-Lie algebra) 
which does not contain the translations. As such, in the (extended) SUSY Lie algebra 
one cannot find a normal sub-Lie algebra complementing to Poincar\'e transformations, 
which obstructs the existence of a homomorphism \emph{onto} the Poincar\'e Lie 
algebra from the (extended) SUSY Lie algebra. Only homomorphic injection 
of the Poincar\'e Lie algebra \emph{into} the (extended) SUSY Lie algebra 
exists, which is just the reverse way.

\section{Concluding remarks}
\label{secconclusions}

A unification mechanism for local gauge and spacetime symmetries was presented. 
The key ingredient is to allow a solvable normal subgroup in the full gauge group, and 
to only require the Levi factor of the full gauge group to be compact, 
not the entire gauge group itself. This relaxed regularity property of allowed gauge groups 
is the minimal requirement for energy non-negativity. 
The solvable extension of the gauge group is seen not to introduce 
new propagating gauge boson degrees of freedom, which would contradict 
present experimental understanding. It is rather seen to be a set of 
inapparent symmetries, representing ``dressing transformations'' for pure one-particle states in a formal 
quantum field theory setting. The unification mechanism also provides an example 
for a non-supersymmetric extension of the group of spacetime symmetries, 
circumventing the McGlinn and Coleman-Mandula no-go theorems in a non-SUSY way. 
Therefore, the construction of invariant Lagrangians to such a local 
unified symmetry group is worth to study. That involves representation theory 
of non-semisimple Lie groups, which is a contemporary branch of research in 
group theory.

\ack

The author would like to thank to William D.\ McGlinn for reading the manuscript, 
and also to M\'aty\'as Domokos and Rich\'ard Rim\'anyi for 
verifying the mathematical validity of the presented results, 
to Zolt\'an Bajnok, P\'eter Vecserny\'es, Zolt\'an Zimbor\'as and Dezs\H{o} Varga for valuable discussions and expert's
opinions on the physical content of the paper as well as for the
discussion on the QFT interpretation. Special thanks to Lars Andersson for 
the kind hospitality in AEI, and to Dimitri Bykov for pointing out an 
interpretational issue concerning O'Raifeartaigh theorem case (iv). 
This work was supported in part by the Momentum (``Lend\"ulet'') program of the
Hungarian Academy of Sciences under grant number LP2013-60, and the J\'anos Bolyai Research
Scholarship of the Hungarian Academy of Sciences.

\appendix

\section{\boldmath Details of the concrete example for the 
$\mathrm{U}(1)$ case}
\label{appu1example}

\begin{figure}[!ht]
\begin{center}
\includegraphics[width=\textwidth]{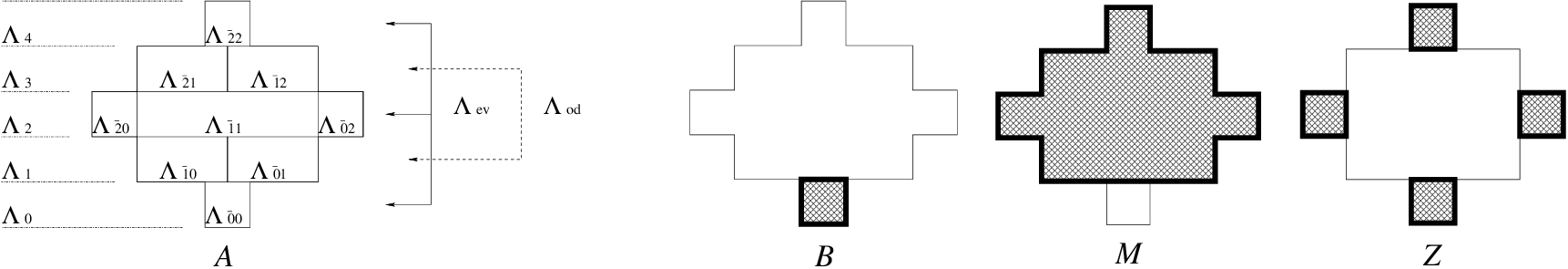}
\end{center}
\caption{Leftmost panel: illustration of the $\Zs$, $\Z$ and $\Zt$ grading 
structure of the spin algebra $A$. 
The unit element $\1$ resides in the subspace $\Lambda_{\bar{0}0}$, whereas the 
canonical generators span the subspace $\Lambda_{\bar{1}0}\oplus\Lambda_{\bar{0}1}$.
Other panels: illustration of the 
important subspaces of the spin algebra, namely the unit subalgebra $B$, 
the maximal ideal $M$, and the center $Z$. 
One unit box depicts one complex dimension on all panels, shaded regions depict 
the subspaces $B$, $M$ and $Z$, respectively.}
\label{figaillustr}
\end{figure}

The spin algebra $A$ has several important linear subspaces. 
Given a canonical generator system $(e_{1},e_{2},e_{1}^{+},e_{2}^{+})$ of $A$, 
the followings can be defined: $\Lambda_{\bar{p}q}$ are the linear 
subspaces of $p,q$-forms, i.e.\ the polynomials consisting of $p$ powers of 
$\{e_{1},e_{2}\}$ and $q$ powers of $\{e_{1}^{+},e_{2}^{+}\}$ ($p,q\in\{0,1,2\}$), 
and one has $A=\mathop{\oplus}\limits_{p,q=0}^{2}\Lambda_{\bar{p}q}$, 
called to be the \emph{$\Zs$-grading} of $A$. Then, there are the linear subspaces of 
$k$-forms, $\Lambda_{k}$, i.e.\ the polynomials consisting of $k$ powers of $\{e_{1},e_{2},e_{1}^{+},e_{2}^{+}\}$ 
($k\in\{0,1,2,3,4\}$), and one has $A=\mathop{\oplus}\limits_{k=0}^{4}\Lambda_{k}$, 
called to be the \emph{$\Z$-grading} of $A$. Finally, there are the subspaces 
$\Lambda_{\ev}$ and $\Lambda_{\od}$ being the even and odd polynomials of $\{e_{1},e_{2},e_{1}^{+},e_{2}^{+}\}$, 
and one has $A=\Lambda_{\ev}\oplus\Lambda_{\od}$, called to be the \emph{$\Zt$-grading} of $A$. 
The subspace $B:=\Lambda_{\bar{0}0}=\C\,\1$ of zero-forms and the subspace 
$M:=\mathop{\oplus}\limits_{k=1}^{4}\Lambda_{k}$ of at-least-1-forms shall play 
an important role as well, and one has $A=B\oplus M$. $B$ is a one-dimensional 
unital associative subalgebra of $A$, spanned by the unity and called the \emph{unit algebra}, whereas $M$ is the so called 
\emph{maximal ideal} of $A$. An other important subspace is 
$Z=\Lambda_{\bar{0}0}\oplus\Lambda_{\bar{2}0}\oplus\Lambda_{\bar{0}2}\oplus\Lambda_{\bar{2}2}$, 
the \emph{center} of $A$, being the largest unital associative subalgebra in $A$ commuting with all elements of $A$. 
All these are illustrated in Figure~\ref{figaillustr}.

In order to study the structure of $\Aut(A)$, 
it is important to note that an element of $\Aut(A)$ maps 
a canonical generator system to a canonical generator system, and that an element 
of $\Aut(A)$ can be uniquely characterized by its group action on an arbitrary 
preferred canonical generator system. Let us take such a system 
$(e_{1},e_{2},e_{1}^{+},e_{2}^{+})$, with occasional notation $e_{3}=e_{1}^{+}$, $e_{4}=e_{2}^{+}$. 
The group structure of $\Aut(A)$ can then be characterized with the following four subgroups:
\begin{enumerate}[(i)]
\item Let $\Aut_{\Zs}(A)$ be the group of $\Zs$-grading preserving 
automorphisms: they act on the canonical generators as 
$e_{i}\mapsto\sum_{j=1}^{2}\alpha_{ij}e_{j}$ and $e_{i}^{+}\mapsto\sum_{j=1}^{2}\bar{\alpha}_{ij}e_{j}^{+}$ ($i\in\{1,2\}$), 
the bar $\bar{(\cdot)}$ meaning complex conjugation and the $2\times 2$ complex matrix $\left(\alpha_{ij}\right)_{i,j\in\{1,2\}}$ being invertible. 
\item Let $\mathcal{J}:=\{I,J\}$ be the two element subgroup of $\Z$-grading 
preserving automorphisms, $I$ being the identity and $J$ being the involutive complex-linear operator of 
\emph{particle-antiparticle label exchanging} acting as $e_{1}\mapsto e_{3}$, $e_{2}\mapsto e_{4}$, $e_{3}\mapsto e_{1}$, $e_{4}\mapsto e_{2}$.
\item Let $\tilde{N}_{\ev}$ be a subgroup of the $\Zt$-grading preserving automorphisms 
defined by the relations $e_{i}\mapsto e_{i}+b_{i}$ and $e_{i}^{+}\mapsto e_{i}^{+}+b_{i}^{+}$ 
with uniquely determined parameters $b_{i}\in\Lambda_{\bar{1}2}$ ($i\in\{1,2\}$).
\item Let $\InAut(A)$ be the subgroup of inner automorphisms, i.e.\ the ones of the form 
$\exp(a)(\cdot)\exp(a)^{-1}$ with some $a\in\Real(A)$. These are of the form 
$e_{i}\mapsto e_{i}+[a,e_{i}]+\frac{1}{2}[a,[a,e_{i}]]$ ($i\in\{1,2,3,4\}$) with 
uniquely determined parameter $a\in\Real(\Lambda_{\bar{1}0}\oplus\Lambda_{\bar{0}1}\oplus\Lambda_{\bar{1}1}\oplus\Lambda_{\bar{2}1}\oplus\Lambda_{\bar{1}2})$.
\end{enumerate}
With these, the semi-direct product splitting
\begin{eqnarray}
\Aut(A) & = & \underbrace{\InAut(A) \rtimes \tilde{N}_{\ev}}_{=:N} \rtimes \underbrace{\Aut_{\Zs}(A) \rtimes \mathcal{J}}_{=\Aut_{\Z}(A)}
\label{eqautadecomp}
\end{eqnarray}
holds. It is seen that a $\Z$-grading almost determines the underlying $\Zs$-grading: 
only the two-element discrete group of label exchanging transformations $\mathcal{J}$ introduces an ambiguity. 
The subgroup $N$ shall be called the group of \emph{dressing transformations}, 
being a nilpotent normal subgroup of $\Aut(A)$. These transformations are mixing 
higher forms to lower forms, i.e.\ do not preserve the $\Z$ and $\Zt$-grading 
defined by our preferred canonical generator system: they map a system of 
canonical generators like $e_{i}\mapsto e_{i}+\beta_{i}$, the elements 
$\beta_{i}$ residing in the space of at-least-2-forms $M^{2}$ ($i\in\{1,2,3,4\}$), 
deforming the original $\Z$ and $\Zt$-grading to an other one. 
By direct substitution it is seen that the transformations (i)--(iv) indeed define 
independent subgroups of $\Aut(A)$, however the proof of decomposition theorem Eq.(\ref{eqautadecomp}) needs a 
bit more complex mathematical apparatus \cite{Laszlo2017un}. The principle of the 
proof is motivated by \cite{Djokovic1978}, studying the automorphism group of ordinary 
finite dimensional complex Grassmann (exterior) algebras.

By scrutinizing the subgroups, it is seen that the group $\mathcal{J}$ of label exchanging 
transformations has the structure of $\Zt$. On the other hand, one has
\begin{eqnarray}
\Aut_{\Zs}(A)\equiv\GL(2,\C)\equiv\mathrm{U}(1)\times\mathrm{D}(1)\times\mathrm{SL}(2,\C),
\end{eqnarray}
where $\mathrm{D}(1)$ is the dilatation group, i.e.\ $\R^{+}$ with the real multiplication. 
Note that $\mathrm{D}(1)\times\mathrm{SL}(2,\C)$ is nothing but the universal covering 
group of the (proper) homogeneous Weyl group. As far as a fixed 
$\Zs$-grading is taken, $A$ can be always represented via ordinary 
two-spinor calculus, and the algebra identification $A\equiv\Lambda(\bar{S}^{*})\otimes\Lambda(S^{*})$ 
can greatly ease the calculations due to well-known identities in that formalism \cite{PR1984,Wald1984}. 
The group of dressing transformations $N$, however, does not fit automatically 
into that framework: it needs the proper apparatus of the introduced spin 
algebra formalism, or care is needed when represented in terms of two-spinors.

\subsection{Important representations of the example group}
\label{sectreprtheory}

\begin{figure}
\begin{center}
\includegraphics[width=\textwidth]{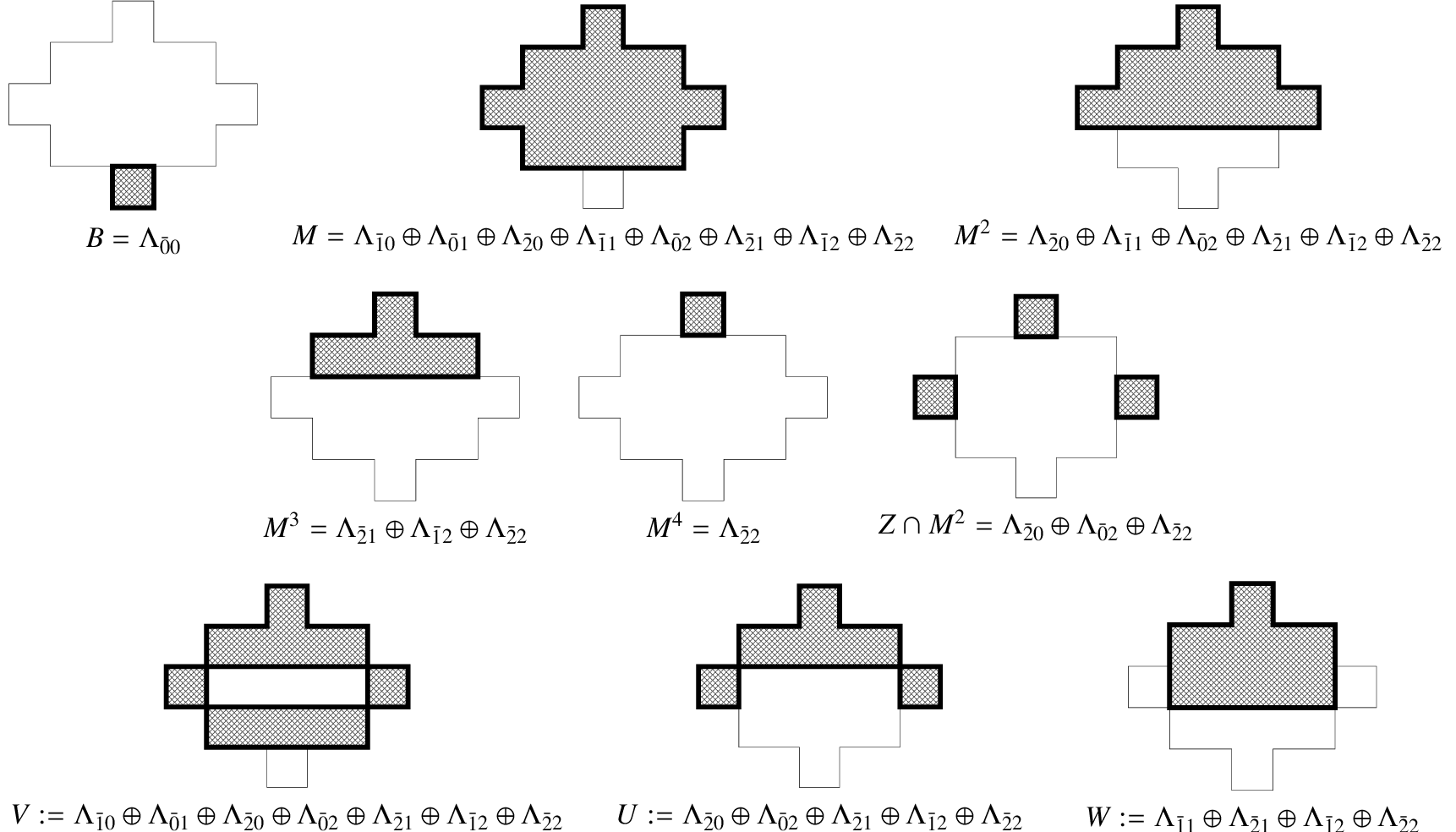}
\end{center}
\caption{Illustration of the $\Aut(A)$-invariant indecomposable subspaces of the 
spin algebra $A$. One unit box depicts one complex dimension, shaded regions 
denote the invariant subspaces on all panels.}
\label{figainvillustr}
\end{figure}

Due to the presence of the nilpotent normal subgroup $N$, $\Aut(A)$ is not 
semisimple. As a consequence, there can be nontrivial invariant subspaces even in the defining 
representation, i.e.\ when $\Aut(A)$ acts on $A$. However, for the same reason, 
the existence of an invariant subspace in a representation of $\Aut(A)$ does not 
imply the existence of an invariant complement. The indecomposable $\Aut(A)$-invariant 
subspaces of $A$ are listed and illustrated in Figure~\ref{figainvillustr}. 
The invariance of these is seen via the orbits of the subspaces 
$\Lambda_{\bar{p}q}$ ($p,q\in\{0,1,2\}$) by the group action of $\mathcal{J}$ and of $N$.

\begin{figure}
\begin{center}
\includegraphics[width=\textwidth]{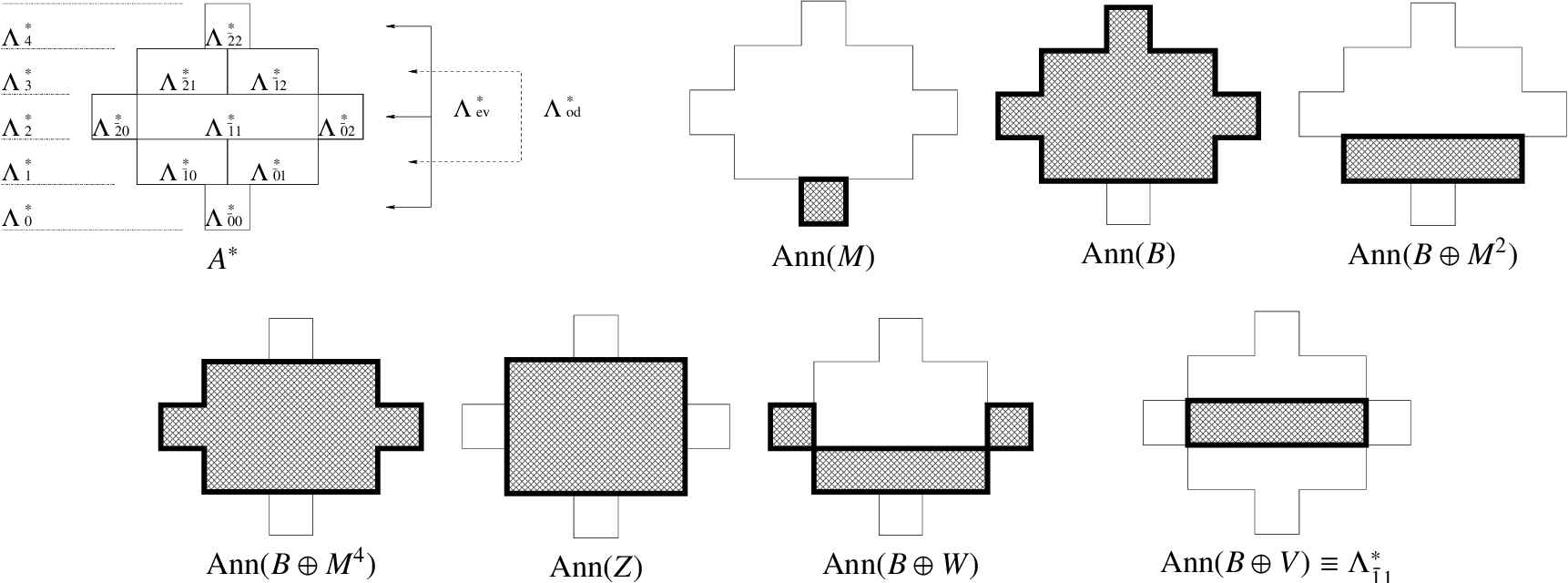}
\end{center}
\caption{Top left panel: illustration of the $\Zs$, $\Z$ and $\Zt$ grading 
structure of the dual vector space $A^{*}$ of the spin algebra $A$. Other panels: 
illustration of the $\Aut(A)$-invariant indecomposable subspaces of the 
dual vector space $A^{*}$ of the spin algebra $A$. One unit box depicts one complex 
dimension, shaded regions denote the invariant subspaces on all panels. Note 
that the subspace $\Ann(B\oplus V)\equiv\Lambda_{\bar{1}1}^{*}$, illustrated on the bottom right panel, 
is a four-vector representation of $\Aut(A)$ and the pertinent group acts there 
as the homogeneous Weyl group.}
\label{figastarinvillustr}
\end{figure}

The group $\Aut(A)$ naturally acts on $A^{*}$, the dual vector space of the 
spin algebra $A$ with the transpose group action. It may be easily seen that the 
$\Aut(A)$-invariant subspaces of $A^{*}$ can be obtained as annulators of 
$\Aut(A)$-invariant subspaces of $A$ itself.\footnote{Given a linear subspace 
$X\subset A$, its annulator subspace $\Ann(X)\subset A^{*}$ is the set of all 
$A^{*}$ elements which maps the subspace $X$ to zero.}
The indecomposable $\Aut(A)$-invariant subspaces of $A^{*}$ are 
listed and illustrated in Figure~\ref{figastarinvillustr}.

In Figure~\ref{figastarinvillustr} it is seen that the 
$\Aut(A)$-invariant subspace
\begin{eqnarray}
\Ann(B\oplus V)\equiv \Lambda_{\bar{1}1}^{*}
\label{eqannboplusv}
\end{eqnarray}
is nothing but a four-vector representation of $\Aut(A)$, on which $\Aut(A)$ acts as 
the homogeneous Weyl group. In the two-spinor representation $A\equiv \Lambda(\bar{S}^{*})\otimes\Lambda(S^{*})$ 
one has simply $\Lambda_{\bar{1}1}^{*}\equiv\bar{S}\otimes S$. The kernel of the corresponding homomorphism 
of $\Aut(A)$ onto the homogeneous Weyl group is said to be the \emph{full gauge group}, 
having the structure $N\rtimes \mathrm{U}(1)$. Given a four dimensional real vector 
space $T$, any injection $T\rightarrow\Real(\Lambda_{\bar{1}1}^{*})$ is called a 
\emph{Pauli injection}, which is the analogue of the ``soldering form'' in the traditional two-spinor calculus 
\cite{PR1984,Wald1984}, extending the group action of $\Aut(A)$ onto the real four dimensional 
vector space $T$. In the usual Penrose abstract index notation that is nothing 
but the usual mapping $\sigma_{a}^{AA'}$ between spacetime vectors $T$ and hermitian 
mixed spinor-tensors $\Real(\bar{S}\otimes S)$. It is seen that the group of dressing transformations $N$ 
respects this basic relation of two-spinor calculus and hence realizes the 
group action of $\Aut(A)$ on the spacetime vectors $T$ as the homogeneous Weyl group.

From Eq.(\ref{eqautadecomp}) it is seen that the connected component $\Aut_{{}_{0}}(A)$ of our concrete example $\Aut(A)$ has the group structure
\newline
\begin{minipage}{\textwidth}
\begin{eqnarray}
\cr
  \underbrace{\underbrace{\underbrace{\overset{\tikzmark{c}}{N}}_{\mathrm{dressing\;transformations}}\rtimes\Big(\underbrace{\overset{\tikzmark{d}}{\mathrm{U}(1)}}_{\mathrm{compact\;internal}}}_{\mathrm{full\;gauge\;(internal)\;group}}\;\times\;\underbrace{\overset{\tikzmark{e}}{\mathrm{D}(1)\times\mathrm{SL}(2,\C)}}_{\mathrm{Weyl\;symmetries}}\Big)}_{\mathrm{symmetries\;of\;}A\mathrm{-valued\;fields\;at\;a\;point\;of\;spacetime\;or\;momentum\;space}}\cr
\tikz[remember picture,overlay] \draw[<-] (c.north) |- +(0,2.5ex) -| (d.north);
\tikz[remember picture,overlay] \draw[<-] (c.north) |- +(0,2.5ex) -| (e.north);
\label{examplestructure}
\end{eqnarray}
\end{minipage}
\newline
which indeed follows the pattern of Eq.(\ref{equnificationmechanism}), providing 
a demonstrative example of the proposed unification mechanism. Again, the 
arrow diagram is meant to indicate that which subgroup acts nontrivially on 
which normal subgroup. Subgroups not connected by arrows do not act on each-other.

\subsection{Optional exclusion of the Weyl dilatations}
\label{secdilatations}

It was seen that $\Aut(A)$ provides a nontrivial unification of the Weyl and 
the $\mathrm{U}(1)$ internal symmetry group. Clearly, $\Aut(A)$ acts on 
the invariant subspace of the maximal forms $M^{4}$ by only a scaling 
due to the dilatation group $\mathrm{D}(1)$. We call the subgroup of 
$\Aut(A)$ acting trivially on the maximal forms $M^{4}$ as \emph{special 
automorphism group of $A$}, and denote it by $\SAut(A)$. By construction, 
the connected component of $\SAut(A)$ has the group structure
\newline
\begin{minipage}{\textwidth}
\begin{eqnarray}
\cr
  \underbrace{\underbrace{\underbrace{\overset{\tikzmark{c}}{N}}_{\mathrm{dressing\;transformations}}\rtimes\Big(\underbrace{\overset{\tikzmark{d}}{\mathrm{U}(1)}}_{\mathrm{compact\;internal}}}_{\mathrm{full\;gauge\;(internal)\;group}}\;\times\;\underbrace{\overset{\tikzmark{e}}{\mathrm{SL}(2,\C)}}_{\mathrm{Lorentz\;symmetries}}\Big)}_{M^{4}\mathrm{-preserving\;symmetries\;of\;}A\mathrm{-valued\;fields\;at\;a\;point\;of\;spacetime\;or\;momentum\;space}}\cr
\tikz[remember picture,overlay] \draw[<-] (c.north) |- +(0,2.5ex) -| (d.north);
\tikz[remember picture,overlay] \draw[<-] (c.north) |- +(0,2.5ex) -| (e.north);
\end{eqnarray}
\end{minipage}
\newline
and is the same as Eq.(\ref{examplestructure}), but without the Weyl dilatations. 
This shows that the inclusion of the subgroup of dilatations is not crucial 
for such unification to happen, but is very natural to include those.

\subsection{Adding the translation or diffeomorphism group}
\label{sectranslations}

Adding translations to the presented homogeneous Weyl (or Lorentz) group 
extension is trivial. One simply takes a four dimensional real affine space 
$\mathcal{M}$ as the model of the flat spacetime manifold, with underlying 
vector space (``tangent space'') $T$. One takes in addition the 
spin algebra $A$, and constructs the trivial vector bundle 
$\mathcal{M}\times A$. The algebraic product on $A$ extends to the sections 
of this vector bundle (i.e.\ to the $A$-valued fields) pointwise, being translationally invariant. 
Given a Pauli injection (soldering form) between $T$ and $\Real(\Lambda_{\bar{1}1}^{*})$, 
$\Aut(A)$ acts on $T$ as the homogeneous Weyl group (or $\SAut(A)$ acts on 
$T$ as the homogeneous Lorentz group). The vector bundle 
automorphisms of $\mathcal{M}\times A$ preserving the algebraic product of 
fields as well as preserving the Pauli injection shall have the desired group 
structure including both the spacetime translations and $\Aut(A)$ in a 
semi-direct product:
\newline
\begin{minipage}{\textwidth}
\begin{eqnarray}
\cr
\overset{\tikzmark{f}}{\mathcal{T}}\;\rtimes\;\overset{\tikzmark{g}}{\Aut_{{}_{0}}(A)} = \cr
\cr
\cr
\cr
  \underbrace{\Big(\underbrace{\overset{\tikzmark{a}}{\mathcal{T}}}_{\mathrm{translations}}\times\underbrace{\underbrace{\overset{\tikzmark{c}}{N}}_{\mathrm{dressing\;transformations}}\Big)\rtimes\Big(\underbrace{\overset{\tikzmark{d}}{\mathrm{U}(1)}}_{\mathrm{compact\;internal}}}_{\mathrm{full\;gauge\;(internal)\;group}}\;\times\;\underbrace{\overset{\tikzmark{e}\;\;\tikzmark{b}}{\mathrm{D}(1)\times\mathrm{SL}(2,\C)}}_{\mathrm{spacetime\;related}}\Big)}_{\mathrm{global\;symmetries\;of\;}A\mathrm{-valued\;fields\;when\;considered\;over\;flat\;spacetime}}\cr
\tikz[remember picture,overlay] \draw[<-] (a.north) |- +(0ex,4.5ex) -| (b.north);
\tikz[remember picture,overlay] \draw[<-] (c.north) |- +(0ex,2.5ex) -| (d.north);
\tikz[remember picture,overlay] \draw[<-] (c.north) |- +(0ex,2.5ex) -| (e.north);
\tikz[remember picture,overlay] \draw[<-] (f.north) |- +(0ex,2.5ex) -| (g.north);
\label{globalautha}
\end{eqnarray}
\end{minipage}
\newline
as a global symmetry of fields, following the pattern of Eq.(\ref{equnifiedglobal}). When acting on $\mathcal{M}$, it shall act 
as the Poincar\'e group combined with global metric rescalings. 
This also implies a causal structure on $\mathcal{M}$. Clearly, Eq.(\ref{globalautha}) 
is a non-supersymmetric extension of the Poincar\'e group, circumventing 
McGlinn and Coleman-Mandula no-go theorems. As noted previously, using 
$\SAut(A)$ the whole construction can be performed also without including 
the metric dilatations.

The ``gauging'' of $\Aut(A)$, i.e.\ making $\Aut(A)$ (or $\SAut(A)$) a local symmetry is also trivial. 
Let $\mathcal{M}$ be a four dimensional real manifold modeling the spacetime 
manifold, with tangent bundle $T(\mathcal{M})$. Take in addition a vector bundle 
$A(\mathcal{M})$ whose fiber in each point is spin algebra. Take also a 
pointwise Pauli injection between $T(\mathcal{M})$ and $\Real(\Lambda_{\bar{1}1}^{*})(\mathcal{M})$. 
The gauged version of $\Aut(A)$ shall be nothing but the product 
preserving vector bundle automorphisms of $A(\mathcal{M})$, and they 
act on $T(\mathcal{M})$ as the combined group of diffeomorphisms and pointwise 
spacetime metric conformal rescalings, being the symmetries of (conformal) GR.

\subsection{Meaning of dressing transformations}
\label{secdressing}

In the presented example the physical meaning of the nilpotent normal subgroup 
$N$ can be understood as the ``dressing'' of pure one-particle 
states of a formal QFT model at a fixed spacetime point or momentum. Note, that spin algebra differs from a CAR algebra 
of QFT with the fact that the antiparticle creation operators are not yet 
identified with particle annihilation operators. It can be shown however \cite{Laszlo2017un}, 
that an $\Aut(A)$-covariant family of self-dual CAR algebras can be associated 
to the spin algebra $A$, and vice-versa. Here, the self-dual CAR algebra is a 
mathematical structure, introduced by Araki \cite{Araki1970}, formally 
describing the algebraic behavior of quantum field operators. With the use of 
this relation, the spin algebra is a convenient reparametrization 
of the quantum field algebra of a QFT at a fixed point of spacetime or 
momentum space, revealing the hidden internal symmetry subgroup $N$. The details 
of the spin algebra $\leftrightarrow$ self-dual CAR algebra family 
correspondence is, however, out of the scope of the present paper 
mainly focusing on unification, and shall be rather discussed in \cite{Laszlo2017un}.

\section*{References}
\bibliographystyle{iopart-num}
\bibliography{laszlojphysa}

\end{document}